\documentclass[10pt,a4paper]{article}
\usepackage{amsfonts,amssymb}
\usepackage[dvips]{graphicx}
\DeclareGraphicsExtensions{.eps}
\usepackage{epsfig}
\usepackage{graphicx}
\font\twelvei = cmmi10 scaled\magstep1
       \font\teni = cmmi10 
\font\mbf = cmmib10 scaled\magstep1
       \font\mbfs = cmmib10 \font\mbfss = cmmib10 scaled 833
\font\msybf = cmbsy10 scaled\magstep1
       \font\msybfs = cmbsy10 \font\msybfss = cmbsy10 scaled 833
       \def\mit{\fam1 }
       \def\bmit{\fam9 }
\def\lsim{\raise0.3ex\hbox{$<$}\kern-0.75em{\lower0.65ex\hbox{$\sim$}}}
\def\gsim{\raise0.3ex\hbox{$>$}\kern-0.75em{\lower0.65ex\hbox{$\sim$}}}
\title
{Hydrodynamic simulations of viscous accretion flows around black holes}
\author
{Kinsuk Giri\thanks{kinsuk@bose.res.in}$^{1}$, Sandip K. Chakrabarti
\thanks{chakraba@bose.res.in}$^{1,2}$\\
$^{1}$S. N. Bose National Centre for Basic Sciences, Salt Lake, Kolkata 700098, India\\
$^{2}$Indian Centre for Space Physics, Chalantika 43, Garia Station Rd., Kolkata, 700084, India}

\begin{document}

\date{}


\maketitle

\label{firstpage}

\begin{abstract}
We study the time evolution of a rotating, axisymmetric, viscous accretion flow around black 
holes using a grid based finite difference method. We use the Shakura-Sunyaev 
viscosity prescription. However, we compare with the results obtained when all the three 
independent components of the viscous stress are kept. We show that the centrifugal pressure supported
shocks became weaker with the inclusion of viscosity. The shock is formed farther out when the viscosity 
is increased. When the viscosity is above a critical value, the
shock disappears altogether and the flow becomes subsonic and Keplerian everywhere
except in a region close to the horizon, where it remains supersonic. We also find that as the 
viscosity is increased, the amount of outflowing matter in the wind is decreased to less than a
percentage of the inflow matter. Since the post-shock region could act as a reservoir
of hot electrons or the so-called `Compton cloud', the size of which changes with viscosity,
the spectral properties are expected to depend on viscosity strongly: the harder states 
are dominated by low-angular momentum and the low viscosity flow with a significant outflows
while the softer states are dominated by the high viscosity Keplerian flow having very little outflows.
\end{abstract}


\section{Introduction}

Pringle (1981) pointed out that in present of viscosity, most of the matter of the disc
accretes into the black hole while most of the angular momentum is taken farther
away by very little matter. Though this basic principle is the key to 
believe that the standard models of Shakura \& Sunyaev (1973, hereafter SS73) 
and Novikov \& Thorne (1973) are possible, there are very few numerical simulations which actually 
show how a Keplerian disc is formed, or whether it should always form. 
There are clearly two possibilities: when the companion
supplies high angular matter (as in a low mass X-ray binary), viscous stress must be very large
to transport this in order that the accretion may occur. On the other hand, when mainly winds from
the companion (as in a high mass X-ray binary) is accreted 
the flow need not produce a Keplerian disc (Chakrabarti \& Titarchuk, 1995, hereafter CT95;
Smith, Heindl and Swank, 2002) since the viscosity may not be high enough to 
transport angular momentum rapidly to redistribute the angular momentum. It has been 
shown by Chakrabarti (1989) that low angular momentum, non dissipative flows produce 
axisymmetric standing shock waves in tens of Schwarzschild radii, but the presence of 
a large viscosity (Chakrabarti, 1990, hereafter C90) will remove the shock wave since the Rankine-Hugoniot
relation is not be satisfied in a highly dissipative flow. A confirmation of such an
assertion, originally made in the context of the isothermal flows, came both through numerical simulations
(Chakrabarti \& Molteni, 1995, hereafter CM95) as well as theoretical studies of flows with more general 
equation of states (Chakrabarti, 1996a, hereafter C96; Chakrabarti \& Das, 2004; Das \& Chakrabarti, 2004).
The theoretical works showed that the topology of the flow is modified when viscosity is
added and beyond a critical viscosity, the centrifugal pressure supported shocks do not form.
The numerical work of CM95, which was based on the smoothed particle
hydrodynamics (SPH) in a one dimensional flow, clearly demonstrated the transport
of angular momentum by the viscosity. It showed that for low enough viscosity parameter $\alpha$
(SS73), the standing shock continues to form. However, with the increase in
viscosity, the shock weakens and moves outward. Ultimately, for high enough viscosity the 
shock disappears. Most interestingly, the post-shock subsonic flow acquires a Keplerian distribution.
Thus, for high enough viscosity, the entire disc becomes subsonic and Keplerian, 
the assumed condition for the Shakura-Sunyaev disc.
{Igumenshchev et al. (1996) studied two-dimensional flows but concentrated only the inner region of the 
disk, namely, the region less than $20 r_g$. The main interest was to study the transonic nature of the flow
just before the matter enters into the horizon. They find that a torus like structure 
forms close to the black hole and the angular momentum increases outward.
Lanzafame et al. (1998) carried out the SPH simulation in a region with a radial extent 
of $50 r_g$ in two dimensions and concentrated on the shock formation. In particular, they showed 
that high viscosity can remove the shock waves from the accretion flows. However, 
SPH method is often intrinsically dissipative and it was necessary to verify the 
behaviour with other simulation methods. Igumenshchev et al. (1998) used the finite difference
method and allowed the heat generated by viscosity heating to be radiated away or absorbed totally.
The computational box was up to $300r_g$, but the outer boundary condition was that of a near Keplerian 
flow having no radial velocity. The inner boundary was kept at $3r_g$. Thus the possibility 
of having a shock or the inner sonic point were excluded. The disc was found to be stable
for very high viscosity parameter $\alpha$ and less stable for lower $\alpha$. Igumenshchev et al. (2000)
extended earlier work by studying the dependence on the polytropic index $\gamma$ which varied from
$4/3$ to $5/3$ as well as viscosity parameter $\alpha$ and again found that the stability of the solutions 
depend on these parameters. }

So far, however, the fate of the solutions which include shocks have not been studied. The formation of 
shock required injection of very lower angular and low viscosity flows and the inner boundary
must be inside $2r_g$ in order that the inner sonic point is formed (C89). In Giri et al. 
(2010, hereafter Paper I), the results of standing and oscillating shock formations in inviscid flows 
was presented. We now extend this study by adding viscosity. The theoretical work discussed in 
C89 or Chakrabarti (1996a) were carried out for a one dimensional flow since
they required the sonic point analysis. For a two dimensional flow, a completely self-consistent
theoretical solution is not possible. This is why a numerical simulation is necessary to 
answer the following questions: (a) Are the conclusions based on theoretical considerations
continue to remain valid for a two dimensional flow? (b) Do the shocks survive for 
higher viscosity? (c) How does the outflow rate depend on viscosity? (d) Are 
all the components of the viscous stress important in a thick accretion flow? 
(e) Under what condition does the Keplerian flow form? 

In the next Section, we show how we introduce the viscosity into the grid-based finite
difference method (Molteni, Ryu \& Chakrabarti, 1996). We consider both the cases, namely,
when only the $r\phi$ component is present, and when all the components, namely, $r\phi$,
$\phi z$ and $rz$ components are present. In \S 3 \& 4, we present the methodology and 
the results of our simulation respectively. Finally, in \S 5, we draw concluding remarks. 

\section{Model Equations with Viscosity}

In what follows, we consider a two dimensional axisymmetric flow around a Schwarzschild black hole. 
Instead of using general relativity, we use the well known pseudo-Newtonian potential prescribed by 
Paczy\'nski and Wiita (1980). We use the cylindrical polar coordinates ($r$, $\phi$ and $z$). 
The mass, momentum and energy conservation equations in a compact form using non-dimensional 
units are given in Molteni, Ryu \& Chakrabarti 
(1996; hereafter MRC96). We use the mass of the black hole
$M_{BH}$ (assumed to be ten solar mass in this paper), 
the velocity of light $c$ and the Schwarzschild radius $r_g=2GM/c^2$ as the 
units of the mass, velocity and distance respectively.

The equations governing the inviscid flow have been presented in Ryu et al. (1995),
MRC96 and Paper I in great detail and we do not repeat everything here. In the conservative 
form, the equations are given by,
$$
{\partial{\bmit q}\over\partial t}+{1\over x}{\partial\left(x
{\bmit F}_1\right)\over\partial x}+{\partial{\bmit F}_2\over\partial x}
+{\partial{\bmit G}\over\partial z} = {\bmit S},
\eqno(1a)$$
where the state vector is
$${\bmit q} = \left(\matrix{\rho\cr
                          \rho v_x\cr
                          \rho v_{\theta}\cr
                          \rho v_z\cr
                          E\cr}\right)_, $$
the flux functions are
$${\bmit F}_1 = \left(\matrix{\rho v_x\cr
                         \rho v_x^2\cr
                         \rho v_{\theta}v_x\cr
                         \rho v_z v_x\cr
                         (E+p)v_x\cr}\right)\qquad
{\bmit F}_2 = \left(\matrix{0\cr
                          p\cr
                          0\cr
                          0\cr
                          0\cr}\right)\qquad
{\bmit G} =  \left(\matrix{\rho v_z\cr
                         \rho v_x v_z\cr
                         \rho v_{\theta} v_z\cr
                         \rho v_z^2+p\cr
                         (E+p)v_z\cr}\right)_, \eqno(1b)$$
and the source function is
$$
{\textfont1 = \twelvei
      \scriptfont1 = \twelvei \scriptscriptfont1 = \teni
       \def\mit{\fam1}
{\bmit S} =  \left(\matrix{0\cr
                ~~~\cr
                {\rho v_{\theta}^2\over x}
                -{\rho x\over2\left(\sqrt{x^2+z^2}-1\right)^2\sqrt{x^2+z^2}}\cr
                ~~~\cr
                ~~~\cr
                -{\rho v_x v_{\theta}\over x}\cr
                ~~~\cr
                -{\rho z\over2\left(\sqrt{x^2+z^2}-1\right)^2\sqrt{x^2+z^2}}\cr
                ~~~\cr
                ~~~\cr
                -{\rho \left(xv_x+zv_z\right)\over
                2\left(\sqrt{x^2+z^2}-1\right)^2\sqrt{x^2+z^2}}\cr}\right)_.}
\eqno(1c)$$

Here, energy density $E$ (without potential energy) is defined as,
$E=p/(\gamma-1)+\rho(v_x^2+v_{\theta}^2+v_z^2)/2$, $\rho$ is the mass density,
$\gamma$ is the adiabatic index, $p$ is the pressure, $v_x$, $v_\theta$ and $v_z$
are the radial, azimuthal and vertical component of velocity
respectively. In the case of an axisymmetric flow without 
viscosity, the equation for azimuthal momentum states simply the conservation of
specific angular momentum $\lambda$,
$$
d\lambda/dt=0.
$$

In an inertial frame of reference, the general form of the equations 
of the flow (Batchelor 1967) is
$$
\rho [{\partial {\bf v} \over \partial t} + {{\bf v} . {\nabla {\bf v}}}] = 
- { \nabla P} + {\bf {F_b}}  + {\nabla . {\bf {\tau}}}, \eqno(2)
$$
where, ${\bf v}$ is the flow velocity, $\rho$ is the fluid density, $P$ is the pressure, 
${\bf \tau}$ is the stress tensor, and ${\bf {F_b}}$
represents body forces (per unit volume) acting on the fluid and ${\nabla}$ is the Del operator.
Typically body forces consist of only gravity forces, but may include other
types (such as electromagnetic forces). Here ${\bf {\tau}}$ is
the viscous stress having six mutually 
independent components. In cylindrical coordinates  
the components of the velocity vector given by
${\bf v} = ({v_r}, {v_{\phi}}, {v_z})$.
The six independent components  of the viscous stress tensor (Landau \& Lifshitz 1959)
are listed here in cylindrical coordinates, ${\tau}_{rr} , {\tau}_{r{\phi}} , {\tau}_{rz},
{\tau}_{{\phi}{\phi}}, {\tau}_{{\phi}{z}}$ \& ${\tau}_{zz}.$
If we split all the viscous stress tensor, three components
of  equation $2$ takes the following forms (Landua \& Lifshitz 1959 and Acheson 1990).
The ${v_{r}}$ component of Navier-Stokes equation is given by
$$
\rho [{\partial {v_{r}} \over \partial t} + {v_r} {\partial {v_{r}} \over \partial r}
 + {{{v_{\phi}^2}} \over r} + {{v_{\phi}} \over r}{\partial {v_{r}} \over \partial {\phi}} 
 + {{v_z}{\partial {v_{r}} \over \partial z}}] =
$$
$$
 -{\partial P \over \partial {r}}
  + {\mu} [{{{\partial ^ 2}{v_{r}}} \over {{\partial r} ^ 2}} +
 {{1 \over r} {\partial {v_{r}} \over \partial r}} -{{v_{r}} \over  {r^2}} +
{{1 \over {r^2}} {{{\partial ^ 2}{v_{r}}} \over {{\partial {\phi}} ^ 2}}} +
 {{{\partial ^ 2}{v_{r}}} \over {{\partial z} ^ 2}} -
 {{2 \over {r^2}}{\partial {v_{\phi}} \over \partial {\phi}}}] + {F_{r}}.       \eqno(3a)
$$

Again, the ${v_{\phi}}$ component is given by
$$
\rho [{\partial {v_{\phi}} \over \partial t} + {v_r} {\partial {v_{\phi}} \over \partial r}
 + {{{v_{\phi}}{v_r}} \over r} + {{v_{\phi}} \over r}{\partial {v_{\phi}} \over \partial {\phi}} 
 + {{v_z}{\partial {v_{\phi}} \over \partial z}}] =
$$
$$
 -{{1 \over r} {\partial P \over \partial {\phi}}}
  + {\mu} [{{{\partial ^ 2}{v_{\phi}}} \over {{\partial r} ^ 2}} +
 {{1 \over r} {\partial {v_{\phi}} \over \partial r}} -{{v_{\phi}} \over  {r^2}} +
{{1 \over {r^2}} {{{\partial ^ 2}{v_{\phi}}} \over {{\partial {\phi}} ^ 2}}} +
 {{{\partial ^ 2}{v_{\phi}}} \over {{\partial z} ^ 2}} +
 {{2 \over {r^2}}{\partial {v_{\phi}} \over \partial {\phi}}}] + {F_{\phi}}.   \eqno(3b)
$$

Finally, the ${v_{z}}$ component is given by
$$
\rho [{\partial {v_{z}} \over \partial t} + {v_r} {\partial {v_{z}} \over \partial r}
 + {{v_{\phi}} \over r}{\partial {v_{z}} \over \partial {\phi}} 
 + {{v_z}{\partial {v_{z}} \over \partial z}}] =
$$
$$
 -{\partial P \over \partial {z}}
  + {\mu} [{{{\partial ^ 2}{v_{z}}} \over {{\partial r} ^ 2}} +
 {{1 \over r} {\partial {v_{z}} \over \partial r}} +
{{1 \over {r^2}} {{{\partial ^ 2}{v_{z}}} \over {{\partial {\phi}} ^ 2}}} +
 {{{\partial ^ 2}{v_{z}}} \over {{\partial z} ^ 2}}] + {F_{z}},   \eqno(3c)
$$
where, $\mu$ is the dynamic viscosity defined by $ \mu = {\eta} {\rho} $ and $\eta$ is called
the kinematic viscosity. Here, ${F_{r}},{F_{\phi}}$ \& ${F_{z}}$ are the so-called body forces.
The only body force which is present in our system is the gravitational force.
Thus, $F_{\phi} = {\rho {G_{\phi}}}, F_{r} = {\rho {G_{r}}} \& F_{z} = {\rho {G_{z}}}$, where 
$G_\phi, G_r$ and $G_z$ are the components of acceleration due to gravity, namely,  
$$ G_r = -{1 \over {2 {(R-1)^2}}}{{r} \over R},$$
and
$$ G_z = -{1 \over {2 {(R-1)^2}}}{{z} \over R},$$
where, $ R = \sqrt{r^2+z^2}$. For the axisymmetric case, $G_{\phi} = 0$ 
and thus, $F_{\phi} = 0$. 
As we have chosen the axisymmetric case, we have neglected $\partial \over \partial {\phi} $ added terms. 
So the above equations reduces as following.

Equation $(3a)$ reduces to
$$
\rho [{\partial {v_{r}} \over \partial t} + {v_r} {\partial {v_{r}} \over \partial r}
  + {{v_z}{\partial {v_{r}} \over \partial z}}] =
$$
$$
 -{\partial P \over \partial {r}}- {{{v_{\phi}^2}} \over r}
  + {\mu} [{{{\partial ^ 2}{v_{r}}} \over {{\partial r} ^ 2}} +
 {{1 \over r} {\partial {v_{r}} \over \partial r}} -{{v_{r}} \over  {r^2}} +
 {{{\partial ^ 2}{v_{r}}} \over {{\partial z} ^ 2}}] +{\rho {G_r}}.       \eqno(4a)
$$

Equation $(3b)$ takes the form,
$$
\rho [{\partial {v_{\phi}} \over \partial t} + {v_r} {\partial {v_{\phi}} \over \partial r}
 + {{v_z}{\partial {v_{\phi}} \over \partial z}}] =
$$
$$
 -{{{v_{\phi}}{v_r}} \over r}
  + {\mu} [{{{\partial ^ 2}{v_{\phi}}} \over {{\partial r} ^ 2}} +
 {{1 \over r} {\partial {v_{\phi}} \over \partial r}} -{{v_{\phi}} \over  {r^2}} +
 {{{\partial ^ 2}{v_{\phi}}} \over {{\partial z} ^ 2}}]  .     \eqno(4b)
$$

Equation $(3c)$ reduces to,
$$
\rho [{\partial {v_{z}} \over \partial t} + {v_r} {\partial {v_{z}} \over \partial r}
 + {{v_z}{\partial {v_{z}} \over \partial z}}] =
$$
$$
 -{\partial P \over \partial {z}}
  + {\mu} [{{{\partial ^ 2}{v_{z}}} \over {{\partial r} ^ 2}} +
 {{1 \over r} {\partial {v_{z}} \over \partial r}} +
 {{{\partial ^ 2}{v_{z}}} \over {{\partial z} ^ 2}}] + {\rho {G_z}}.            \eqno(4c)
$$

In case of a thin accretion flow, it is customary to use only
${{\tau}_{r{\phi}}}$  component since it is the dominant
contributor to the viscous stress (SS73). This is responsible for transporting angular momentum along 
the radial direction. The other components are assumed negligible.
Thus, considering only  ${{\tau}_{r{\phi}}}$, the only viscous term which goes into the Eq. $(4b)$ is
$$
{ \mu [{{{\partial ^ 2}{v_{\phi}}} \over {{\partial r} ^ 2}} +
{{1 \over r} {\partial {v_{\phi}} \over \partial r}} -{{v_{\phi}} \over  {r^2}}]} 
$$
$$
= {1 \over {r^2}} {\partial \over \partial r}(r^2{{\tau}_{r \phi}}),   \eqno(5)
$$
where, $${{\tau}_{r \phi}} = {\mu} r {\partial {\Omega} \over \partial r}. $$
Here, $\Omega$ is angular velocity and defined as
$${\Omega} = {{v_{\phi}} \over r}.$$

It has long been suspected that the diffusion of angular momentum through an accretion flow 
is driven by turbulence. The $SS\alpha$-model (SS73) introduced a 
phenomenological shear stress into the equations of motion to model the effects of this 
turbulence. This shear stress is proportional to $p$, the total pressure. This shear
stress permits an exchange of angular momentum between
neighbouring layers. 
Using $SS\alpha$ prescription, we can assume ${\tau}_{r\phi}= -{\alpha p}$,
where $\alpha$ is a proportionality factor which need not be constant throughout the
flow. This has been used in the simulations by CM95 and C96. In Eq. $(5)$ we  put $-{\alpha p}$ in 
the place of ${{\tau}_{r\phi}}$. In that case, the viscous term reduces to
$$
{1 \over {r^2}} {\partial \over \partial r}(-{r^2{\alpha p}})
= -\alpha ({{2 p} \over r} + {\partial p \over \partial r}).   \eqno(6)
$$

 In case of thick accretion flow, all viscous stress could be significant in flow
dynamics. Assuming that the flow is thick $h \sim r$,
the $\alpha$-prescription could be written as (e.g., Igumenshchev et al. 1996) as
$$
\mu = \alpha_s \rho {a^2 \over {{\Omega}_k}}, 
\eqno{(7)}
$$
where, $\alpha_s$ is constant of order 1, $a$ is the adiabatic sound speed, and
$$
 {{\Omega}_k} = [{{1 \over r}{{\partial {\Phi}} \over \partial r}}]^{1 \over 2}
$$
is the Keplerian angular velocity. We carried out simulations using this value 
of $\mu$ and either the $r\phi$ component of the viscous stress or all the 
components of the viscous term. We clearly find major differences in our results.
These will be discussed below. Note that, we used a different notation for the viscosity 
parameter to differentiate between cases where the Shakura-Sunyaev prescription (Eq. 6)
is used or when the definition (7) is used for coefficient of viscosity.

It is to be noted that the heat generated by the viscous dissipation is assumed to
be radiated away instantly. Thus we do not consider any effect of the heating on the 
dynamics of the flow in the present paper.

 In the literature, viscosity prescriptions other than 
those discussed here have also been tried out, 
especially when the shock is present. We followed also the prescription Macfadyen \& 
Woosley (1999) where $\alpha$ was assumed to be constant when $v_{\phi}>v_r$ while it is 
assumed to be scaled as $v_{\phi}/v_r$ in the pre-shock flow to reduce the shear. The result remains
very similar to our present result. This is because the pre-shock flow, where
$v_{\phi}>v_r$, is cooler with a lower thermal pressure and thus the angular momentum
transport rate is weaker. In the post-shock flow, due to high thermal pressure,
the angular momentum transport rate is higher and our disk becomes
similar to Keplerian. Thus a constant $\alpha$ prescription 
plays a role similar to that in a Keplerian disk.

\section{Methodology of the Numerical Simulations}

The setup of our simulation is the same as presented in Paper I.
We consider a viscous, axisymmetric flow in the Pseudo-Newtonian gravitational 
field of a point mass $M_{bh}$ located at the centre in cylindrical
coordinates  $[x,\theta,z]$. We assume that at infinity, the gas pressure is
negligible and the energy per unit mass vanishes. We also assume that
the gravitational field of the black hole can be described by Paczy\'{n}ski \& Wiita (1980),
$$
\phi(r,z) = -{GM_{bh}\over2(R-R_g)}, 
$$
where, $R=\sqrt{r^2+z^2}$ and the Schwarzschild radius is given by
$$R_g=2GM_{bh}/c^2.$$

We also assume a polytropic equation of state for the
accreting (or, outflowing) matter, $P=K \rho^{\gamma}$, where,
$P$ and $\rho$ are the isotropic pressure and the matter density
respectively, $\gamma$ is the adiabatic index (assumed 
to be constant throughout the flow, and is related to the
polytropic index $n$ by $\gamma = 1 + 1/n$) and $K$ is related
to the specific entropy of the flow $s$. In all our simulations, we take $\gamma = 4/3$.

The computational box occupies one quadrant of the r-z plane with $0 \leq r \leq 50$ 
and $0 \leq z \leq 50$. The incoming gas enters the box through the outer boundary 
 (which is, in reality, of a vertical cylinder in shape),
located at $x_b = 50$. We have chosen the density of the incoming
gas ${\rho}_{in} = 1$ for convenience since, in the absence of self gravity
and cooling, the density is scaled out, rendering the simulation results valid for any accretion rate.
We supply the radial velocity $v_r$, the sound speed $a$ (i.e., temperature) 
of the flow at the boundary points. We take the boundary values of density from the standard
vertical equilibrium solution (Chakrabarti, 1989). We inject matter at the outer
boundary along the z-axis. We consider only the first
quadrant and use the reflection boundary condition on the equatorial plane
and z-axis to obtain the solution in other quadrants. 
In order to mimic the horizon of the black hole at one
Schwarzschild radius, we placed an absorbing inner boundary at $R = 1.1 r_g$, 
inside which all material is completely absorbed into the black hole. For the 
background material, we consider of density ${\rho}_{bg} = 10^{-6}$ 
having the sound speed (or, temperature) to be same as that of the incoming
matter. Hence, the incoming matter has a pressure $10^6$
times larger than that of the background matter. This matter is put to avoid
unphysical singularities caused by `division by zero'. This is washed out
and replaced by the incoming matter withing a single dynamical time scale. 
The calculations were performed with $512 \times 512$ cells, so
each grid has a size of $0.097$ in units of the Schwarzschild radius.  The 
timescale of matter from the outer boundary is about $0.07$s as
computed from the sum of $dr/<v_r>$ over the entire radial grid, 
$<v_r>$ being averaged over $20$ vertical grids.

 All the simulations have been carried out assuming 
a stellar mass black hole $(M = 10{M_\odot})$. 
We carry out the simulations for several hundreds of dynamical 
time-scales. In reality, our simulation time corresponds to a few seconds in physical units.
The conversion of out time unit to physical unit is $2GM/c^3$, and thus 
the physical time for which the programme was run would scale with the mass of the black hole.

\section{Simulation Results}

The total variation diminishing (TVD) method that we use here was primarily developed
to deal with fluid dynamics (Harten, 1983). In the astrophysical context, Ryu et al. (1994 \& 1996)
developed the TVD scheme to study astrophysical flows around black holes. 
In Paper I, the 
oscillation phenomena in accretion flows around black holes 
related to the QPO were reported. That study of an inviscid accretion flow 
around black hole showed that the shock location changes with the
change of specific angular momentum $(\lambda)$ and specific energy $(\cal E)$,
both of which were constant. In the present situation of a viscous flow, none of these
is constant. Recently, one dimensional solution for quasi-spherical viscous flow 
was investigated by Lee, Ryu \& Chattopadhyay (2011) who found that there are oscillatory
propagating shocks moving outward when the viscosity is large enough.

\subsection{Isothermal injection at the outer boundary}
As mentioned earlier, we chose the outer boundary of the simulation grid
at $r = 50$. The specific angular momentum ($\lambda$) of the flow is 
chosen to be $1.66$  (For comparison, we note that the marginally stable angular momentum
is $1.83$ in this unit.) and the specific energy ($\cal E$) of the flow at 
the equatorial plane ($z = 0$) is chosen to be $0.0035$. 
The radial velocity pointing to the origin is chosen to be constant in all heights $ v =
{({v_r}^2 + {v_z}^2)^{1 \over 2}} = 0.072$. These injection parameters correspond to 
those of low angular momentum transonic flow solutions and are qualitatively different from those 
of earlier workers such as Igumenshchev et al. (1998, 2000) since they mainly inject near Keplerian flow 
with no radial velocity. From the energy of the flow, we obtain the sound speed at the
equatorial plane to be $0.059$. We employ an isothermal outer boundary condition (MRC96). In other words,
we take the same sound speed at all heights at the outer boundary. We add the 
$SS{\alpha}$ viscous term in the non-viscous system as discussed earlier.
We stop the simulation at $t=24.75$ seconds. This is more than three hundred times the
dynamical time. Thus, the solution has most certainly come out of the transient regime and started exhibiting
solutions characteristics of its flow parameters. The simulation results will be discussed now.

In Figs. 1a and 1b, we compare the Mach number and the specific angular momentum 
variations in the equatorial plane of the flow for various $\alpha$. 
To make the comparison meaningful, all the  runs were carried out up to $t=24.75$s.
Each result is obtained starting with an inviscid flow ${\alpha = 0}$ 
(marked) and then gradually increasing ${\alpha}$ till the shock goes out of the 
grid and eventually disappears. The values of ${\alpha}$ for which 
the curves are drawn are (left to right in Fig. 1a and bottom to 
top in Fig. 1b):  $0.0$, $0.0175$, $0.035$, $0.0525$, $0.06125$, 
$0.07$, $0.0735$ respectively. The shock location shifts outward
with viscosity exactly as predicted in C90 and C96. For an inviscid flow ($\alpha=0$), 
the matter bounces back from the centrifugal barrier and flows outward near the 
equatorial plane. As viscosity is enhanced, the angular momentum is transported outward and
hence specific angular goes up. The velocity goes down and density goes up. This can be seen in 
Fig. 1b in the $3-30\ r_g$ region. In this particular example, the shock disappears 
above $\sim \alpha = 0.074$, which is the critical value of ${\alpha}$ here. Since the grid boundary 
is at a finite distance in a numerical grid it is difficult of show the disappearance of the 
shock since that would require changing the boundary condition dynamically to let the 
shock through when it reached the boundary. The results are indeed similar to the results of 
CM95 where one dimensional viscous isothermal flow was treated using Smoothed 
Particle Hydrodynamics (SPH) code. As described in C90, the critical $\alpha$ ($\alpha_c$) 
is defined by that specific $\alpha$ 
for which the subsonic branch of the transonic flow solution passes through both the inner and
outer sonic points. Thus $\alpha_c$ clearly depends on the injected flow parameters. 
Hence, for a different choice of injected parameters, the critical value of ${\alpha}$ will be different.
\begin{figure}
\includegraphics[height=8.5truecm,width=8.5truecm,angle=0]{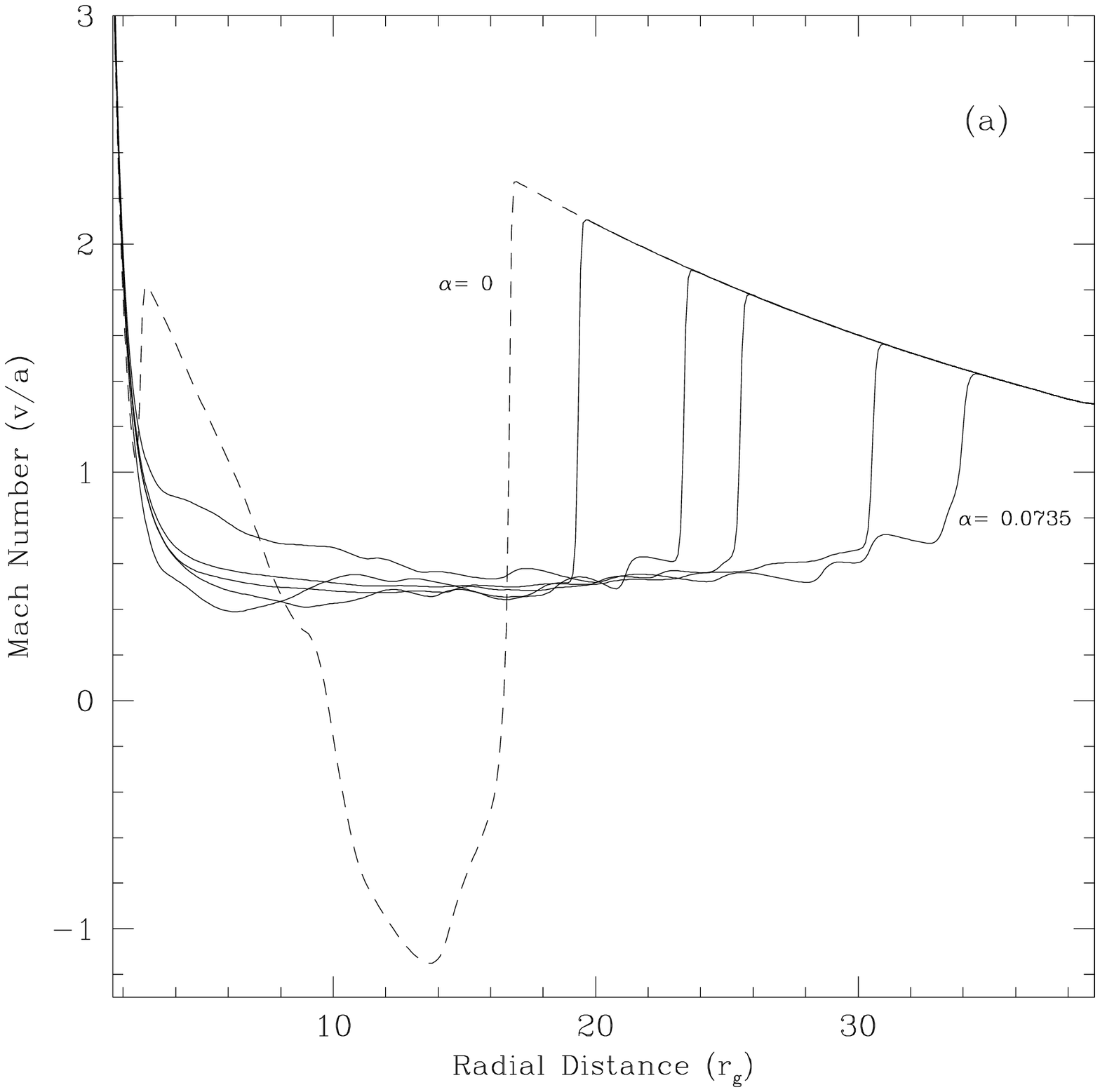}
\includegraphics[height=8.5truecm,width=8.5truecm,angle=0]{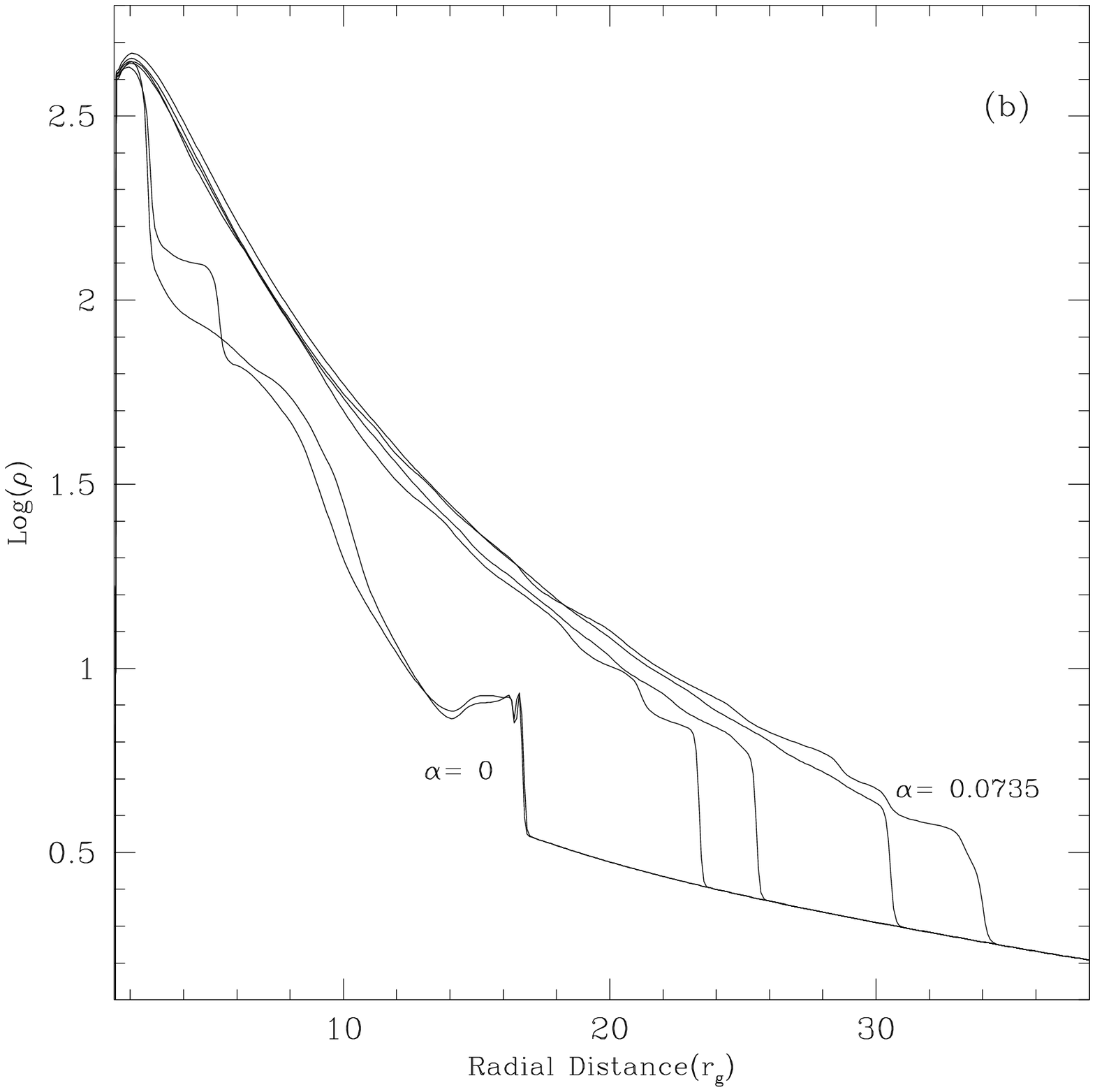}
\caption{Variation of (a) Mach number and (b) density with radial distance in a viscous 
transonic flow on the equatorial plane. As the viscosity parameter is increased, 
the angular momentum is transported outward shifting along with it 
the centrifugal pressure supported shock wave. The $\alpha$ parameters are
[(left to right in (a) and bottom to top in (b)]:  
$0.0$, $0.0175$, $0.035$, $0.0525$, $0.06125$, $0.07$, $0.0735$ respectively. }
\end{figure} 

In Fig. 2, we show how the density of matter and the velocities vary with viscosity. 
We superpose contours of constant density and velocity. The length of the arrows are proportional to velocity,
the longest being that of $v=0.6$. 
The results are for $t = 24.75$s. ${\alpha}=0.0$ 
(a: top left), $0.0525$ (b: top right), $0.0735$ (c: bottom left) $0.074$ (d: bottom right) 
respectively. The contour having minimum density has ${{\rho}_{min}} = 0.2$. 
The maximum density is (a) ${{\rho}_{max}} = 450$, (b) ${{\rho}_{max}} = 395$, 
(c) ${{\rho}_{max}} = 354$ and (d) ${{\rho}_{max}} = 286$. The contour interval is ${\delta\rho} = 0.25$.
We observe that the standing shock is truly two dimensional and oblique 
(Chakrabarti, 1996b).   Here three shocks meet at a point, is a
good example of a prominent triple-shock which forms away from the equatorial plane.
However, it is not a pure triple-shock, since the matter flows from right to the left for 
both the shocks facing the upstream.
As the matter flows in from the right hand side and passes through the shock, it becomes hot and puffs
up as a thick accretion disc (see also, Molteni, Lanzafame and Chakrabarti, 1994, hereafter MLC94).
The standing shock moves outwards as the viscosity parameter is increased. For ${\alpha} = 0.074$ which is 
larger than the critical ${\alpha}$ for this case and the shock disappears. In a numerical
simulation with a finite injection radius the shock, albeit very weak, stays very close
to the outer boundary. The piled up angular momentum in the post-shock flow clearly drives the shock outwards.
\begin{figure}
\begin{center}
\includegraphics[height=14truecm,width=14truecm,angle=0]{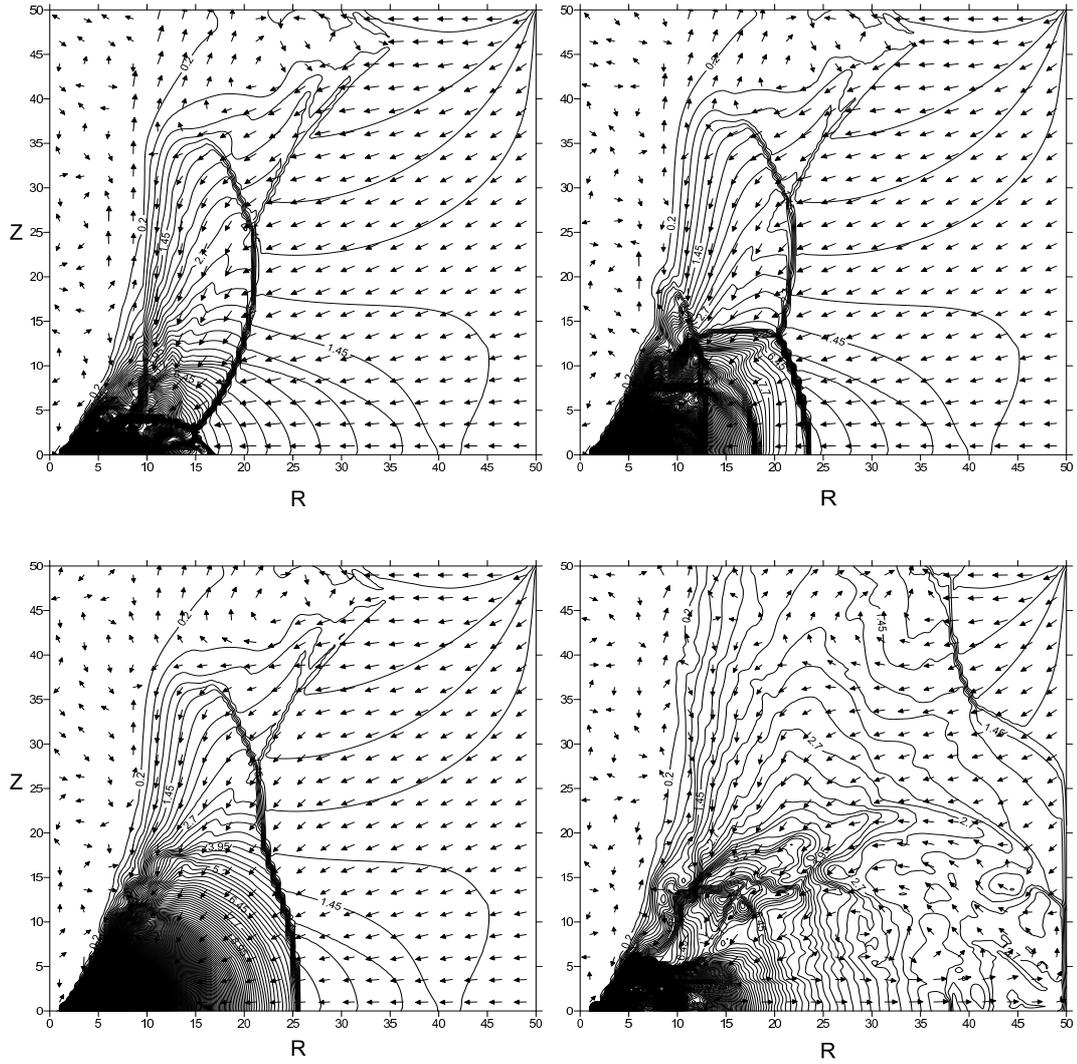}
\caption{Changes in the density and velocity distribution with the change of viscous
parameter ${\alpha}$ at $t= 24.75$s. Here, densities
are in normalized unit, radius and velocity are in Schwarzschild unit. Here,
${\alpha}=0.0$ (top left), $0.0525$ (top right), $0.0735$ (bottom left)
$0.074$  (bottom right)respectively. For details see the text.}
\end{center}
\end{figure}
In Fig. 3 we plot the time variation of the shock location at the equatorial plane 
for various viscous parameter ${\alpha}$. The values of ${\alpha}$ are (from bottom to top):
$ 0.0$, $ 0.0175$, $ 0.035$, $ 0.0525$, $0.06125$, $ 0.0735$ respectively. It is clear that the
average shock locations are shifting outwards when the value of ${\alpha}$ increase. 
Shocks close to the black hole exhibit lower amplitude and higher frequency oscillations 
while those farther out show opposite effects.  This is mainly because the frequency is
decided by the inverse of the infall time in the post-shock region. The compression wave in the 
post-shock region bounces back from the centrifugal barrier and pushes the shock outward.
At some point the outward journey is stopped when post-shock pressure drops and the shock
turns back. Most interestingly, for $\alpha =  0.0525$ 
and $\alpha = 0.06125$, the oscillation of the shock disappears and 
standing shocks are formed while for other $\alpha$s there are oscillations. 
This is not surprising, since, as was shown for the inviscid flow (Chakrabarti, 1989)
as well as the viscous flow Chakrabarti \& Das (2004), the Rankine-Hugoniot 
relations are satisfied only in a limited region of the parameter space, and beyond 
the critical viscosity the relation not satisfied at all. In this case,
the flow has two sonic points and the high entropy solution must pass through the 
inner sonic point (C89). Hence, the flow generates entropy through a shock jump,
but the location itself cannot be fixed because the Rankine-Hugoniot condition is 
not satisfied. This creates an unstable situation and consequently,
the shocks can oscillate (see, Ryu, Molteni \& Chakrabarti, 1996; Paper I).
\begin{figure}
\begin{center}
\includegraphics[height=10truecm,width=10truecm,angle=0]{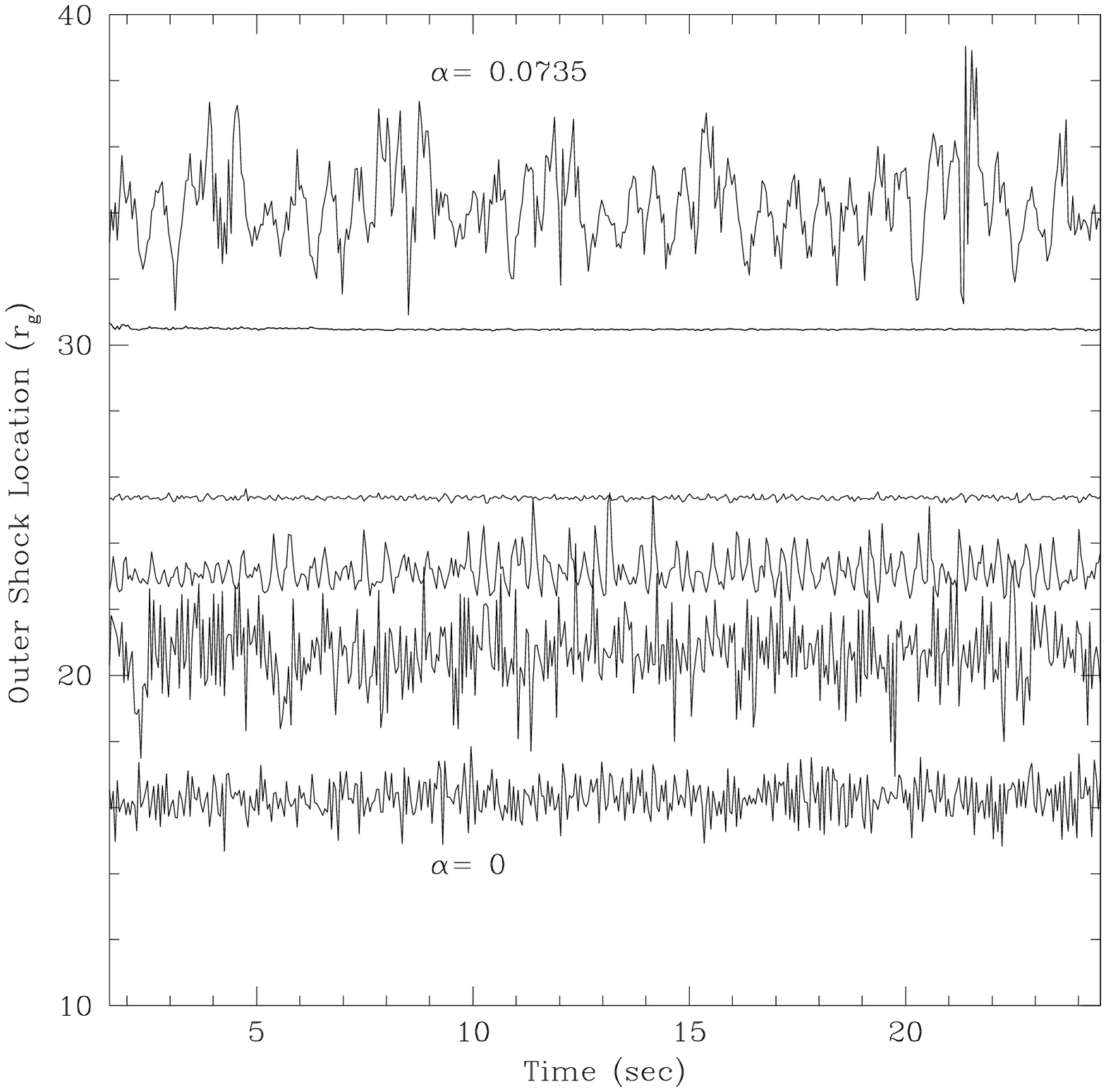}
\caption{Variation of the shock location at the equatorial plane
with time for various viscous parameter ${\alpha}$ (from bottom to top:
$ 0.0$, $ 0.0175$, $ 0.035$, $ 0.0525$, $ 0.06125$, $ 0.0735$ respectively).}
\end{center}
\end{figure} 

\begin{figure}
\begin{center}
\includegraphics[height=10truecm,width=10truecm,angle=0]{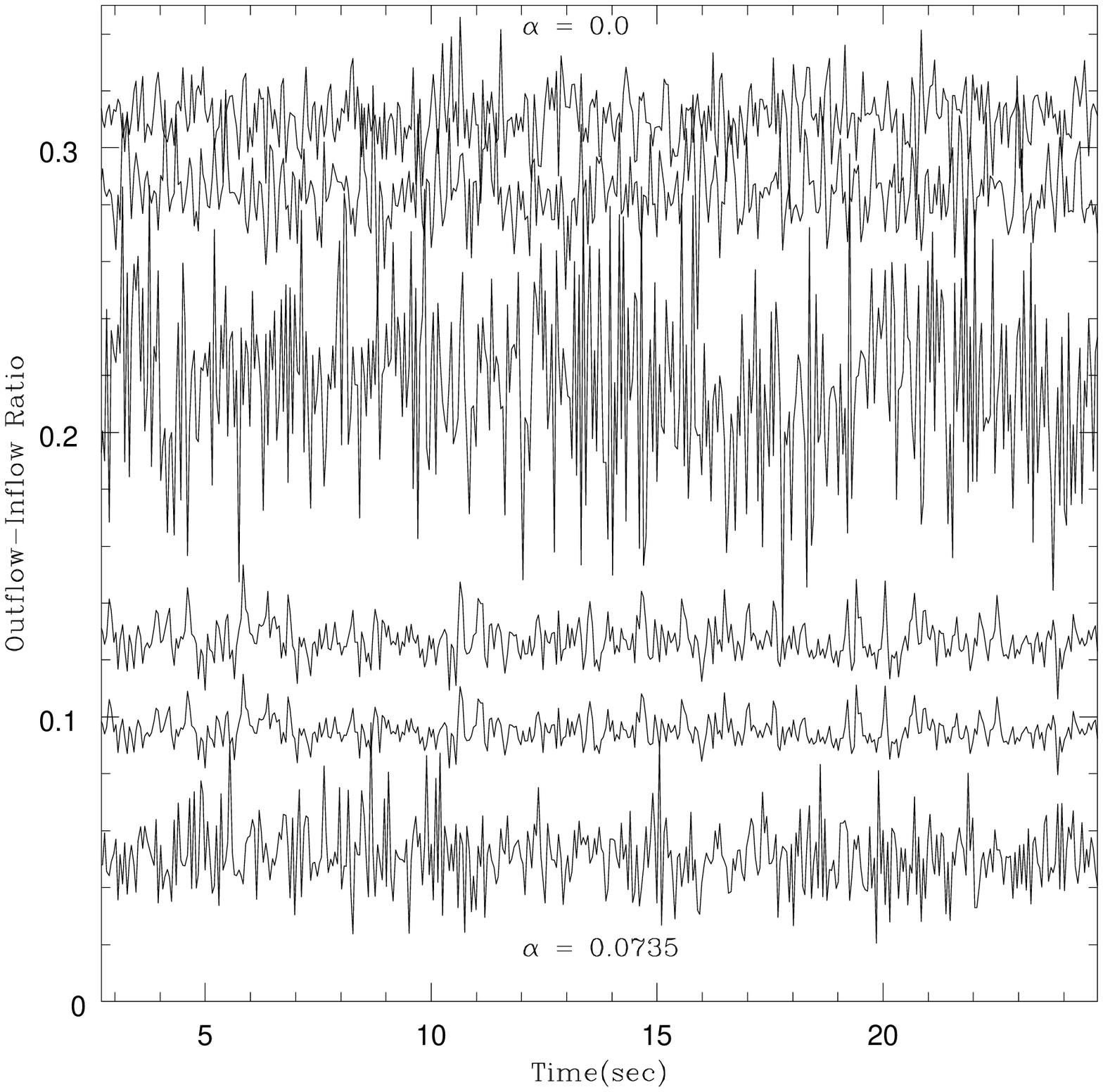}
\caption{ Time variation of the ratio of the outflow to inflow rates
as viscosity parameter is enhanced. Though there are short time scale
fluctuations, the average values decrease as viscosity is increased showing a direct 
relation of the outflow rate with the strength of the shock. The viscosities are
same as in Fig. 3 (from the top curve to the bottom curve).}
\end{center}
\end{figure} 
It is interesting to study the effects of viscosity on the outflow rate.
We observed that with viscosity the shock recedes and becomes weaker.
In Chakrabarti (1999), it was suggested that the ratio $R_{\dot m}$ of the outflow
to the inflow rate would be guided by the compression ratio at the shock. 
In Fig. 4, we plot the time variation of the ratio of the outflow to inflow
rate as the viscosity is enhanced. The values are same as in Fig. 3 (from the top 
curve to the bottom curve). We clearly notice that although the 
ratios exhibit short time scale fluctuations, the average values decrease
as viscosity is enhanced. This could have been guessed also from Fig. 2(a-d),
where the lengths of the outgoing arrows in Figs. 2(a-b) are reduced in number and size in Figs. 2(c-d). 
The longest arrow corresponds to $v\sim0.6$.

\subsection{Vertical equilibrium at the outer boundary}
We now use a different model for injection in order to impress that 
the basic results remain the same even 
when we assume the flow to be in vertical equilibrium (Chakrabarti, 1989) at the outer boundary. 
We inject through $50$ grids out of $512$ grids, i.e., 
the height of the disc at injection is nearly $5$ Schwarzschild radii. 
In this case, the injection rate of the momentum density 
is kept uniform throughout the injected height at the outer edge. 
The specific angular momentum ($\lambda$) and specific energy ({\it \cal E}) 
at the outer boundary is chosen to be the same as in the previous case. 
Here too we stop the simulation at $t = 24.75$ second. The results of the 
simulation are discussed now.

In Figs. 5a and 5b, we show the distribution of Mach number on the equatorial
plane and that {\it averaged} over $15$ grids from the equatorial plane
when $\alpha$ is $ 0.0$, $0.018$, $0.0225$, $0.0315$, $0.05 $ and 
$ 0.09$ respectively (from leftmost curve to the rightmost curve). 
As the $\alpha$-parameter is increased, the flow behaviour changes dramatically.
As in the the previous case, the shock rapidly propagates outward
due to the faster transport of angular momentum in the post-shock flow as compared to the 
pre-shock flow. However, since in the vertical equilibrium model, the 
ram pressure of the injected flow is less, the turbulence and the back 
flow on the equatorial plane remains important even for high $\alpha$. 
\begin{figure}
\includegraphics[height=8.5truecm,width=8.5truecm,angle=0]{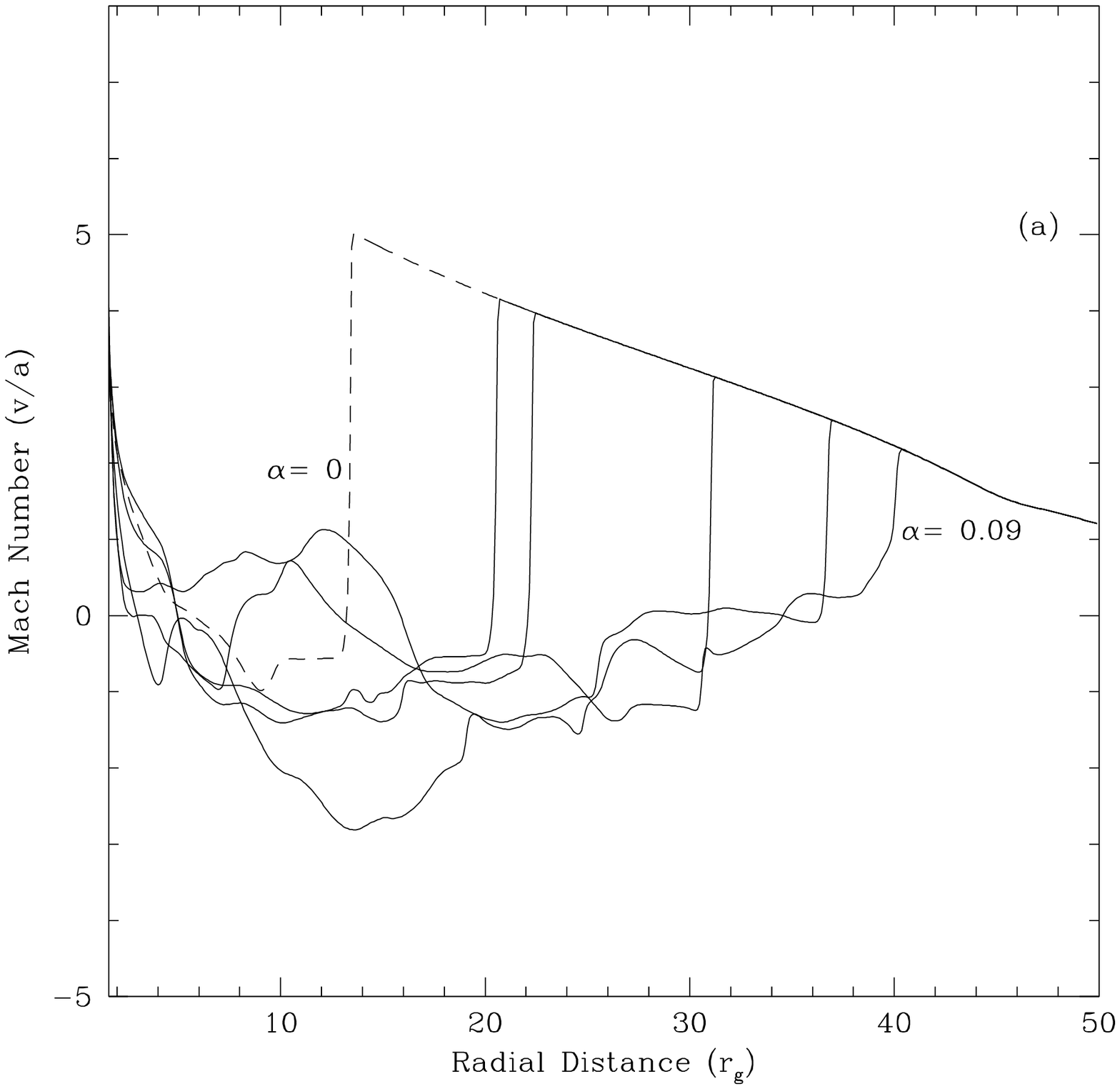}
\includegraphics[height=8.5truecm,width=8.5truecm,angle=0]{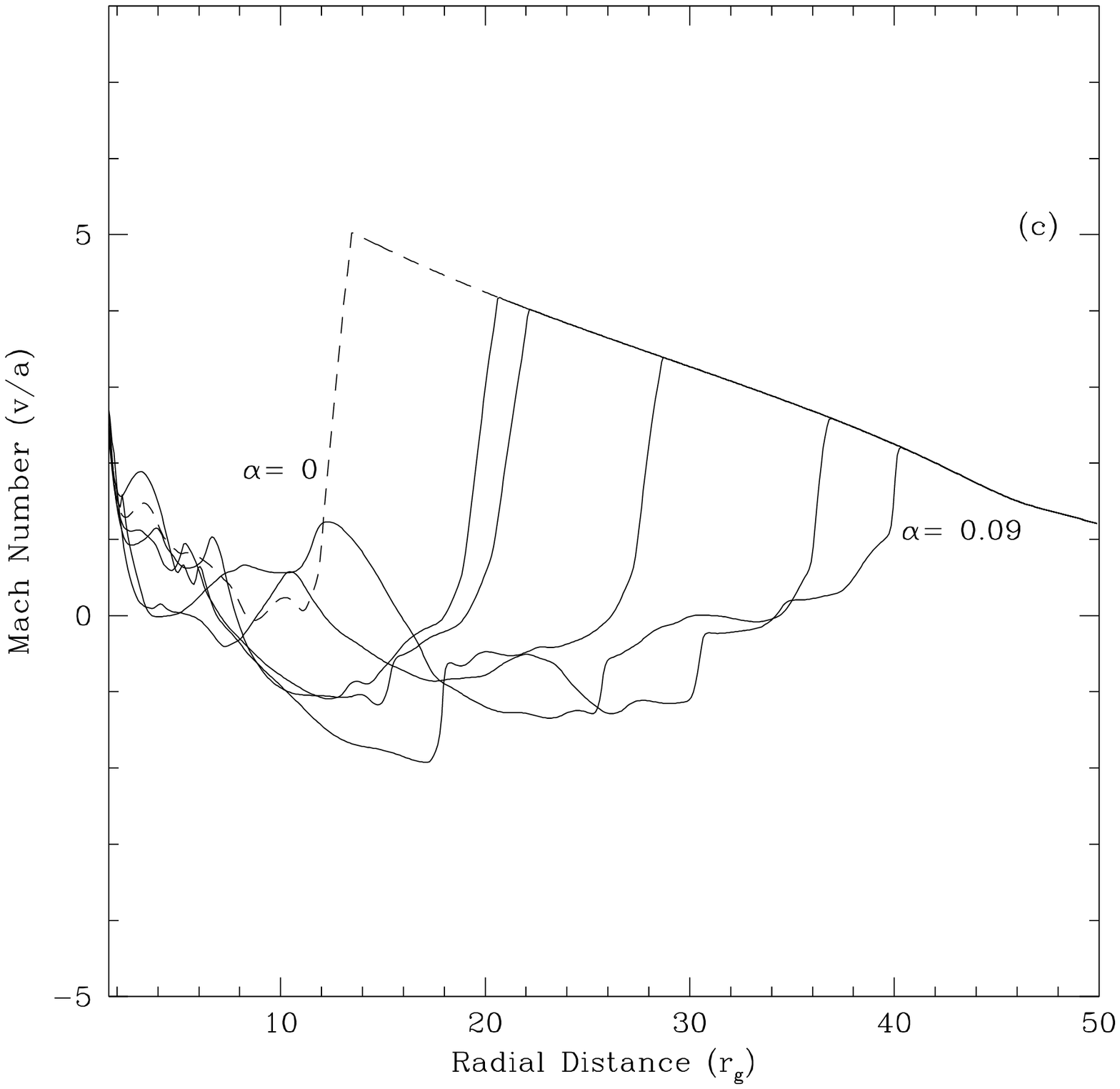}
\caption{Variation of the (a) Mach number on the equatorial plane and (b) the Mach number averaged 
over $15$ vertical grids as a function of the radial distance on the equatorial plane as the $\alpha$
is increased from $0$ (leftmost curve) to $0.09$ (rightmost curve). See the text for details.} 
\end{figure}

Figures 6(a-d) show the distribution of  average specific angular momentum of the flow 
(density weighted average over $15$ vertical grids from the equatorial plane) at the end 
of our simulation for different values of $\alpha$ (thin solid curve). For comparison, we plot 
the specific angular momentum distributions of a Keplerian orbit (thick upper curve). We also plot a `Keplerian' 
distribution (thick lower curve) at a height of $200$ grids ($\sim 20$ Schwarzchild radii) 
and compare the $15$ vertical grid average angular momentum distribution that is obtained from the simulation
at that height (dashed curve). The latter `Keplerian' distribution
was obtained by equating the horizontal component of the gravitational force at that
height with the centrifugal force. The results are shown at $t=24.75$s. The ${\alpha}$-parameters
chosen for the Figs. 6a, 6b, 6c and 6d are 
$0.0$, $0.01$, $0.05$, and $0.09$ respectively. We note that as the viscosity is increased,
the distribution in the post-shock region gradually becomes closer to the 
Keplerian distribution, although, below $r=3$, the distribution is always highly sub-Keplerian.
At the shock, the distribution shows a jump. This is because
for a given $\alpha$, the transport rates in the pre- and the post-shock flows
are different, being very high in the post-shock region due to higher pressure. 
For the high enough viscosity, when the shock reaches infinity (large distance),
the angular momentum distribution becomes that of a Keplerian flow, which is what is expected.
This is thus one scenario by which a Keplerian disc may form in a highly viscous flow.
\begin{figure}
\begin{center}
\includegraphics[height=14truecm,width=14truecm,angle=0]{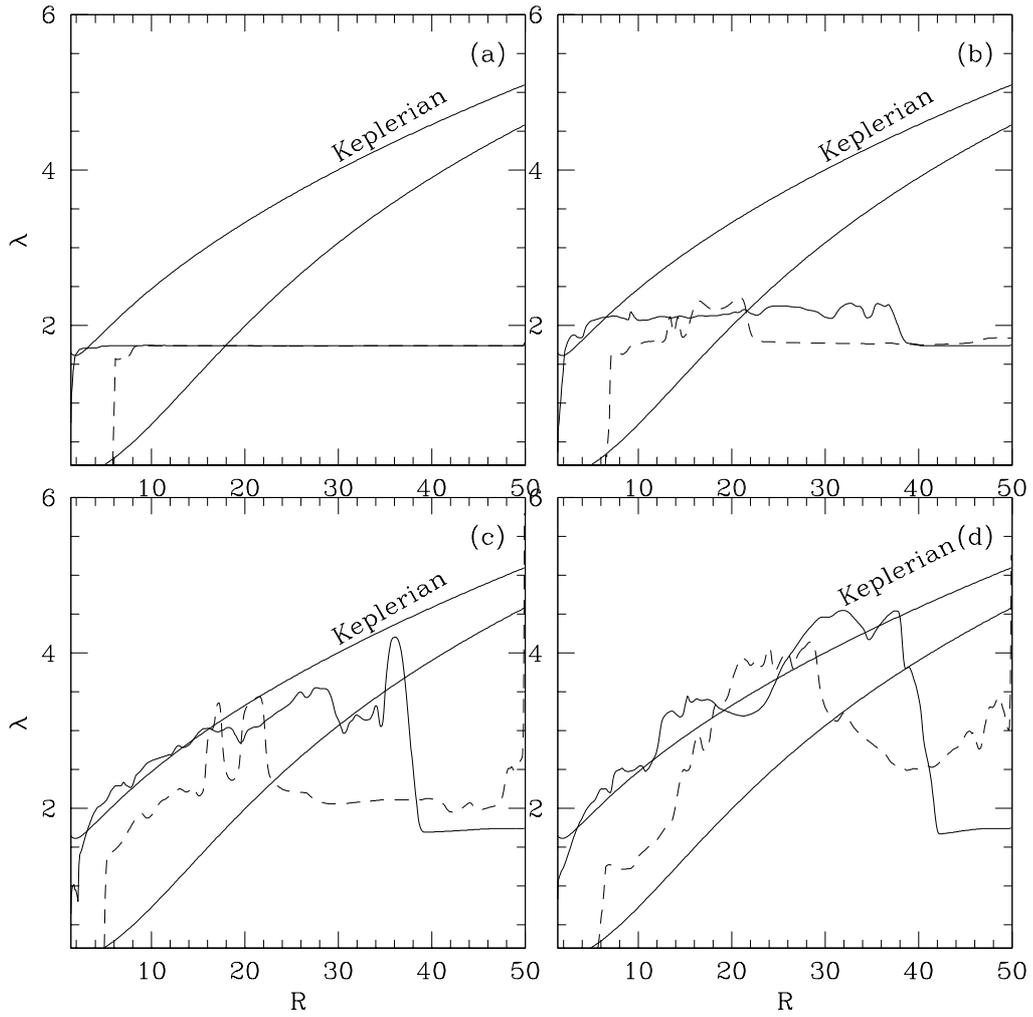}
\caption{Comparison of the specific angular momentum distributions (thin curves) of
 an accretion flow as viscosity parameter $\alpha$ is varied. $\alpha=$ 
(a) $0.0$, (b) $0.01$, (c) $0.05$ and (d) $0.09$. For all the cases, the result is 
compared with the Keplerian angular momentum distribution (thick curves).}
\end{center}
\end{figure}

\subsection{Viscous flow with a constant height injection in radial direction}

For the sake of comparison with the results of inviscid flows presented in Paper I,
we simulate the case where the injection is purely in the $-X$ direction. The injected flow at the 
outer boundary has the same sound speed (temperature) as obtained from the theoretical `constant height'
model. The specific energy and specific angular momentum are  ${\it E} = 0.003$ and ${\lambda} =  1.76$ 
respectively.  The simulation was carried out up to $t=7.63$s, 
or about one hundred dynamical time
(computed as a sum of $dr/<v_r>$ over the whole radial grid, $<v_r>$ being averaged over $20$ 
vertical grids). 
 
Figure 7 shows the distribution of Mach number along the equatorial plane for  $\alpha = 0.0$ (dashed), 
$0.02$ (solid), $0.035$ (long dashed), $0.07$ (dash-dotted) and $0.09$ (log dash-dotted) 
respectively. As in Paper I, we see the formation of a strong (outer) shock at $\sim 25$ 
and a weak (inner) shock at $\sim 3$ when the flow is inviscid. For $\alpha = 0.02$ and 
$0.035$ both the shocks still form, albeit being farther out and weaker.
For higher viscosity, the inner shock completely disappears. The 
outer shock disappears at even higher viscosity. This shows that the critical viscosity
for the inner shock is much lower than that for the outer shock. Since, theoretically only
the outer shock was predicted, this is therefore a totally new result
and could not have been anticipated without numerical simulations.
\begin{figure}
\begin{center}
\includegraphics[height=10truecm,width=10truecm,angle=0]{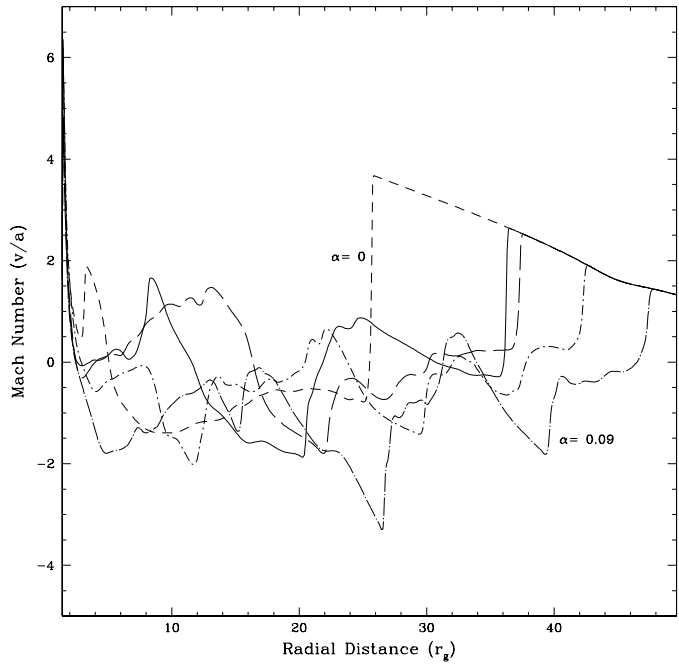}
\caption{Radial distribution of the Mach number on the equatorial plane for a flow with a constant 
height injection in the radial direction at the outer boundary. The viscosity 
parameters are: $\alpha = 0.0$ (dashed), $0.02$ (solid), $0.035$ (long dashed), $0.07$ (dash-dotted) 
and $0.09$ (log dash-dotted). The inner shock disappears at about $\alpha \sim 0.05$ 
and the outer shock disappears at $\alpha \sim 0.1$.}
\end{center}
\end{figure}
In Fig. 8, we show the density and velocity distributions for $\alpha = 0$ (top left) and $0.02$ (top right) 
respectively at the end of the simulation. We zoom the inner region to show how 
the inner shock has shifted from $\sim 3$ (bottom left) to $\sim 8$ 
(bottom right) as viscosity is increased.

\begin{figure}
\begin{center}
\includegraphics[height=14truecm,width=14truecm,angle=0]{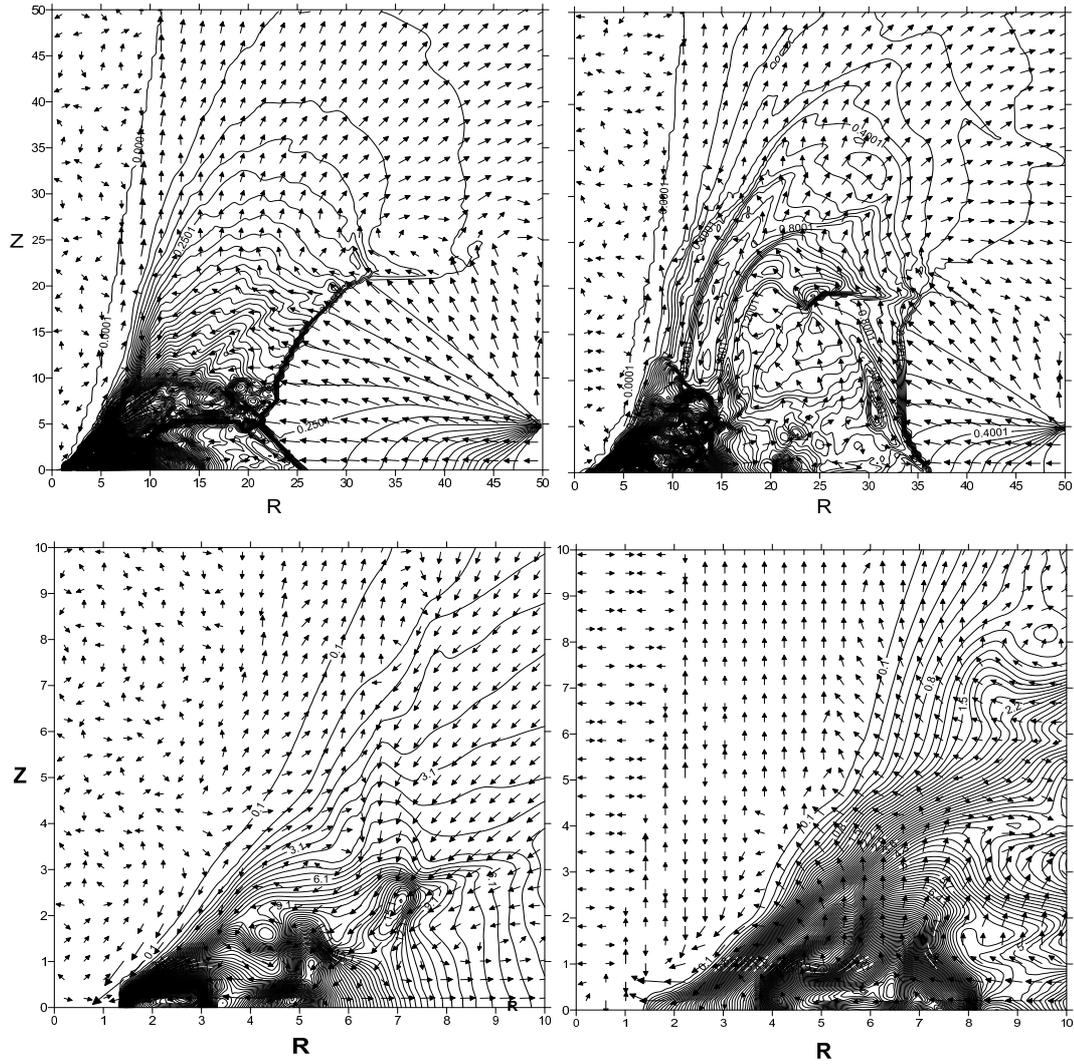}
\caption{The density and velocity distributions for $\alpha = 0$ (top left) and $0.02$ (top right) 
respectively at the end of the simulation where the flow was injected with a 
constant height. We zoom the inner region to show 
the shifting of the inner shock from $\sim 3$ (bottom left) to $\sim 8$ (bottom right) 
when viscosity is increased.}
\end{center}
\end{figure}

\subsection{Effects of boundary location}

 It may be noted that we have run the simulations above for viscosity parameters slightly
below the critical viscosity, since the shock is nearing the outer boundary. As we have 
injection of matter at the outer boundary, it would prevent the shock from leaving the 
grid near the injection area. In order to prove that the shock actually runs away and disappears, 
one requires an infinitely large boundary. However, that this must happen can be easily shown
by running a few cases using the viscosity parameter from both sides of the critical value. 
In Fig. 9(a-b), we present two such cases with (a) ${\cal E}=0.0035$, $\lambda=1.66$, $x_b=50$
and (b) ${\cal E}=0.002$, $\lambda=1.66$, $x_b=100$. The viscosity parameters are 
marked on the curves. The critical viscosities are $\alpha_c\sim 0.0738$ and $0.0325$ respectively.
We clearly see that for $\alpha <\alpha_c$, the shock first goes out farther before 
returning back and settling down at a certain finite distance with some small amplitude oscillation.
However, for $\alpha >\alpha_c$, the shock never returns back and continues to go outward. This
behaviour is independent of the location of the boundary.

\begin{figure}
\begin{center}
\includegraphics[height=10truecm,width=14truecm,angle=0]{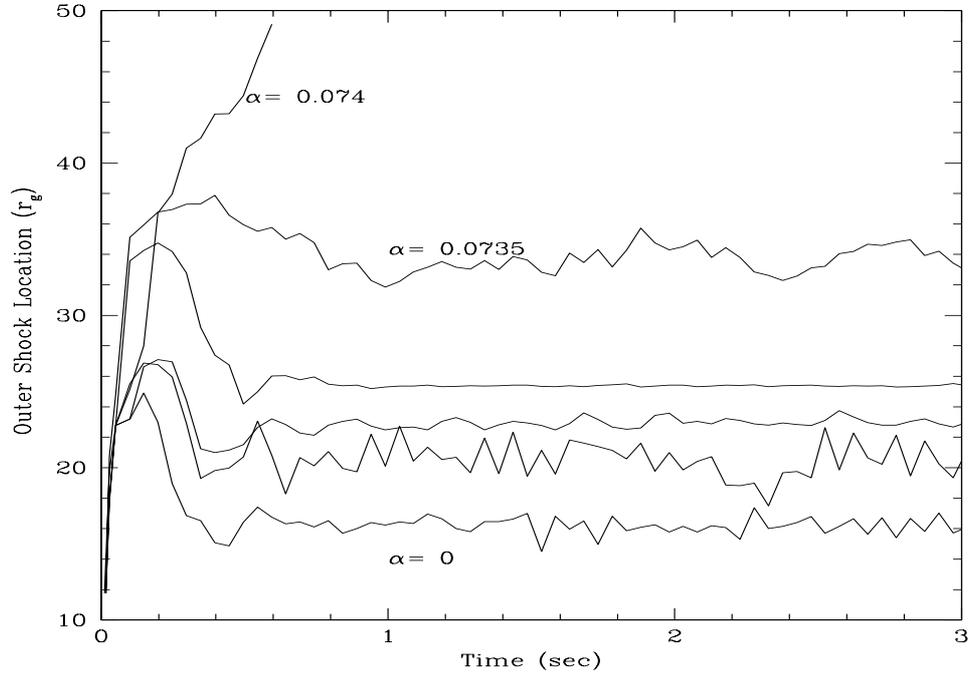}
\includegraphics[height=10truecm,width=14truecm,angle=0]{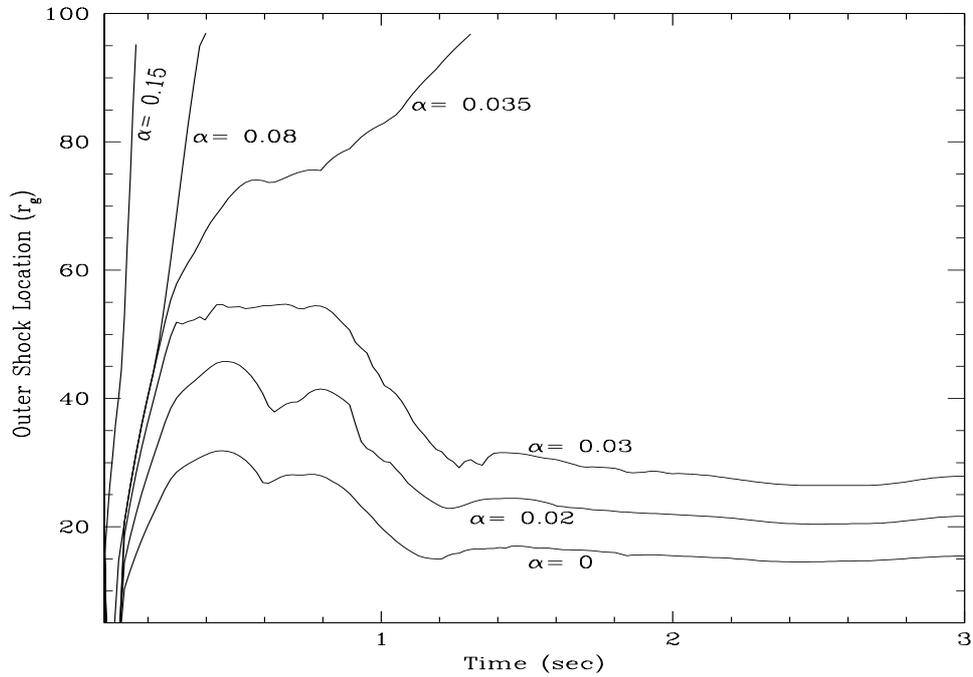}
\caption{Evolution of the shock location as the viscosity parameter $\alpha$ is changed
for two different boundary conditions: (a) $x_b=50, {\cal E} = 0.0035, \lambda = 1.66$;
(b) $x_b=100, {\cal E} =0.002, \lambda=1.66$. The critical viscosities are $\alpha_c=0.0738$ 
and $0.0325$ respectively in these cases. In both the cases, 
when $\alpha <\alpha_c$, the shock first goes out farther before
returning back and settling down, while
for $\alpha >\alpha_c$, the shock never returns back and continues to go outward. }
\end{center}
\end{figure}

\subsection{Effects of viscous stress components}

In this case, we used Eq. (7) in Eqs. 4(a-c). We considered these cases to 
understand the importance of various components of the viscous stress. 
In Fig. 10(a-d), radial distribution of the Mach number and the specific angular
momentum on the equatorial plane and that averaged over fifteen vertical grids from the
equatorial plane are shown. The dashed-dotted curve shows the results when only the $r\phi$ component is used
and the solid curve shows the results when all the three components are used. 
The results are plotted at $t=10$s. The flow parameters are ${\cal E}=0.035, \lambda=1.66$ and $\alpha_s=0.03$.
We also plot the results for the inviscid flows for comparison (dotted curve). Note that
the distribution of angular momentum inside the shock remains almost constant
when a single ($r\phi$) component is used while it becomes similar to Keplerian
when all the three components are included.

\begin{figure}
\begin{center}
\includegraphics[height=14truecm,width=14truecm,angle=0]{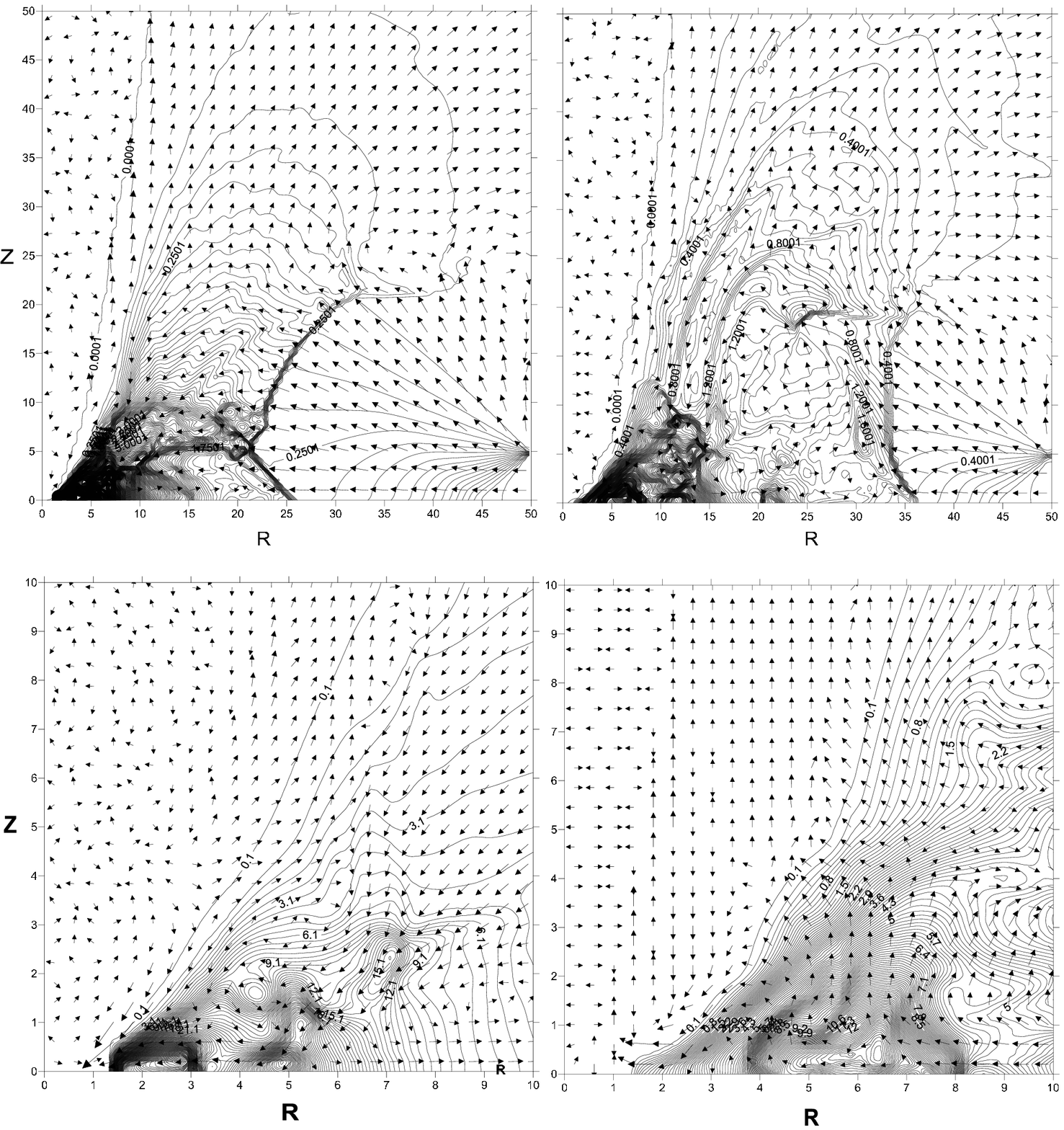}
\caption{Radial distribution of the (a \& b) Mach number, and the (c \& d) specific angular 
momentum on the equatorial plane (a \& c) and that averaged over fifteen vertical grids from the 
equatorial plane (b\& d) for the cases when only the $r\phi$ component (dashed-dotted)
and all the three components (solid) are used. The results are plotted at $t=10$s. 
We also plot the results for the inviscid flow for comparison (dotted). 
The angular momentum inside the shock remains almost constant 
when a single ($r\phi$) component is present while it becomes similar to Keplerian 
when all the three components are included.}
\end{center}
\end{figure}

In Figs. 11(a-d), we plot the distribution of density and velocity when only $r\phi$ component of
the viscous stress is used (top left). Note that the jaggedness of the
shock goes away when all three components of the viscous stress are included (top right). A density maximum
occurs in the  post-shock region. Thus the post-shock region behaves like
a thick accretion disk (Paczy\'nski \& Wiita, 1980) as was also pointed out in SPH simulations
(MLC94). The post-shock region, formed purely due to the centrifugal force, is known as the 
CENtrifugal pressure dominated BOundary Layer or CENBOL. This region is believed to be responsible
to emit hard X-rays in black hole candidates by inverse Comptonizing soft photons
coming from Keplerian disks believed to form near the equatorial plane where the 
viscosity is high (CT95). The corresponding specific angular momentum distributions
are in the bottom-left and the bottom-right panels respectively. Note that up to the disk center
the angular momentum rises rapidly and becomes almost Keplerian as shown in Fig. 11(c-d).
The parameters chosen are the same as in Fig. 10.
A comparison of the behaviour with and without all the three components indicates that the behaviour becomes smoother 
with transport of momentum takes place in all directions. The outflow also has a larger specific angular momentum. 

\begin{figure}
\begin{center}
\includegraphics[height=14truecm,width=14truecm,angle=0]{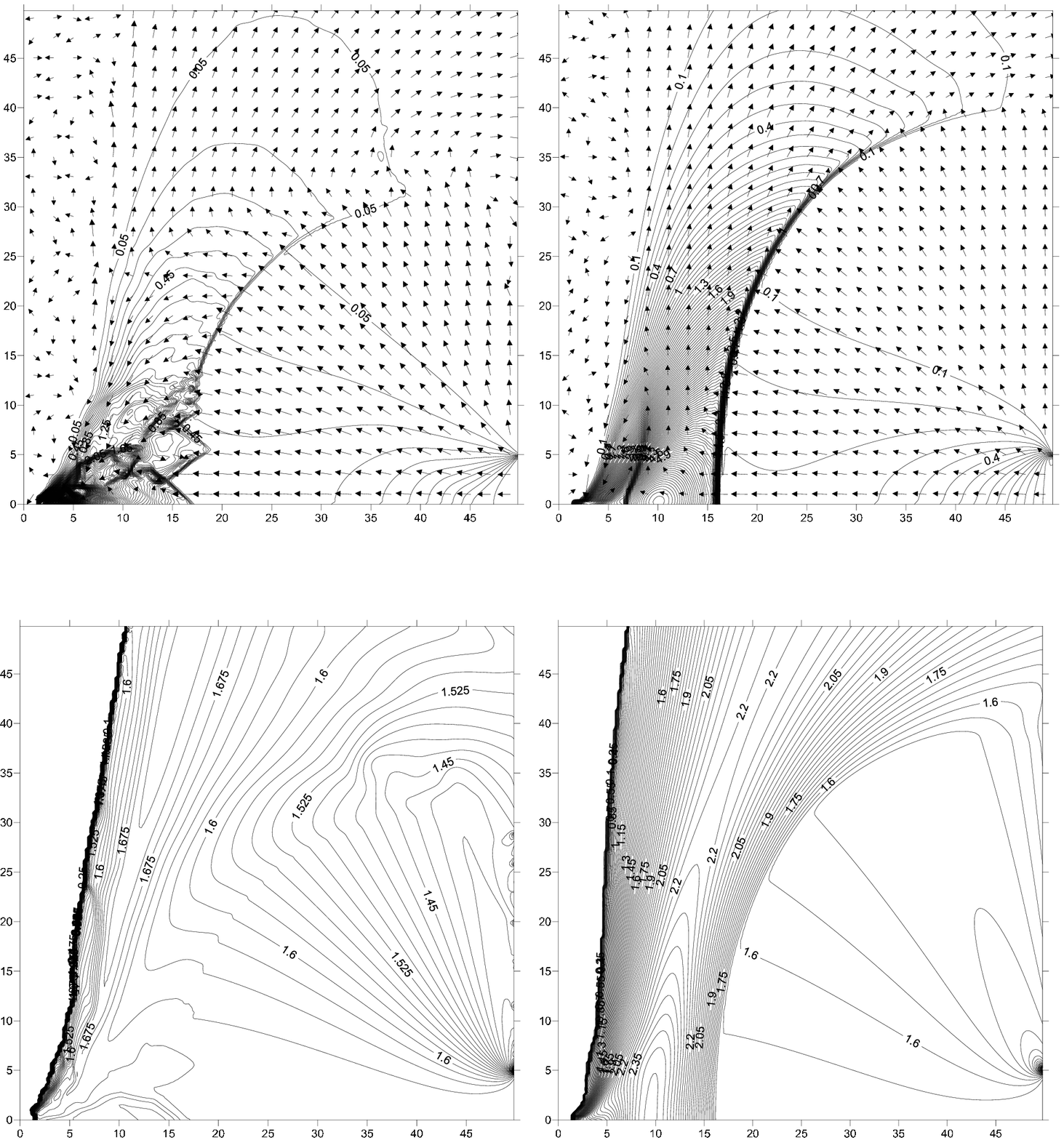}
\caption{Distribution of density and velocity when only $r\phi$ component of 
the viscous stress is used (top left). Note that the jaggedness of the 
shock goes away when all three components are included (top right). Density maximum
and a consequent thick accretion disk like structure is formed in the post-shock
region in the latter case. The corresponding specific angular momentum distributions
are in bottom left and bottom right respectively. 
The parameters are same as in Fig. 10.}
\end{center}
\end{figure}

\section{Discussions}

In this paper, we have presented the results of the numerical simulations of a two dimensional, axisymmetric, viscous
accretion flows. Three parameters, namely, the specific angular momentum, the specific energy
and the viscosity parameter determine the complete solution, although results depend somewhat on the
injection process at the outer boundary. While both inviscid (Chakrabarti, 1989)
and viscous (Chakrabarti \& Das, 2004) flows allow solutions with or without centrifugal barrier
dominated shock waves, we concentrated mostly on the cases when the shocks are formed. We
find that the shocks move outward as the viscosity is enhanced and the post-shock region roughly attains a 
Keplerian distribution. When the viscosity parameter is very high, the shock moves to a large distance 
and the whole disc becomes a Keplerian disc. We also found that the condition of standing shock 
wave formation may be satisfied only in a narrow range of the viscous parameter (keeping other parameters
as constants) which is in line with the conclusions drawn in Chakrabarti \& Das (2004). When the 
Rankine-Hugoniot conditions are not satisfied, the shocks tend to oscillate (Ryu, Chakrabarti,
Molteni, 1997; Paper I) and the frequency of oscillation is decreased and the amplitude is increased
as the shock moves out. 

In contrast to these simulations, the simulations of Igumenshchev and collaborators (1996, 1998, 2000)
concentrated on shock-free solutions. Either the simulation boundary was too small or too big with high
viscosity, the shocks did not form in these simulations. Thus a direct comparison 
is not possible. In some of the models (e.g., CT95) the post-shock
region or CENBOL is treated as the Compton cloud which plays a major role in shaping the 
spectrum of the accretion flows. Similarly, the CENBOL is thought to be responsible 
for the outflows. In Chakrabarti (1999) it was estimated that the ratio of the outflow to the inflow varies with the 
strength of the shock. In the present case, we varied the strength of the shock by increasing viscosity 
and found that the outflow rate, though fluctuating in small time scales, becomes 
smaller as the viscosity parameter is increased.

In Paper I, we emphasized the formation of a weaker inner shock closer to the black hole. 
This was not predicted by the theoretical works. In the present paper, we find that this 
inner shock also becomes weaker and moves outward  when the viscosity is introduced. 
The critical viscosity parameter for removal of the inner shock is lower than what is
needed to remove the outer shock.

In Molteni, Sponholz and Chakrabarti (1996) and  Chakrabarti, Acharyya \& Molteni (2004)
it was shown that the shocks oscillated when the cooling time scale matches with the 
infall time scale. The simulation did not include viscosity. In Ryu, Chakrabarti and Molteni (1997) 
no cooling or viscosity was added, but the shocks were shown to oscillate when the Rankine-Hugoniot relation was
not satisfied. In the present paper, for the first time, we show that even for
a viscous flow, the same conclusion holds true as long as the viscosity is less than the critical value
and the shocks are not totally removed.

It is well known that the light curves of the radiation coming from an accretion disc
around a black hole exhibit quasi-periodic oscillations (QPOs). It appears that in the absence of a hard surface,
the standing shock itself behaves like a hard surface and its oscillation changes the 
size of the post-shock region significantly. This could be the cause of the 
low and intermediate frequency quasi-periodic oscillations (Chakrabarti \& Manickam, 2000; Rodriguez et al. 2004;  
Remillard et al. 2006; Gliozzi et al. 2010; Qu et al. 2010). When the viscosity 
is increased, the Keplerian rate is enhanced and at the same time, the shock recedes to 
a large distance and the time period is increased. Asymptotically, this means that a Keplerian disc should not show 
QPOs. We also observe that away from the equatorial plane the angular momentum is sub-Keplerian
(i.e., smaller compared to the specific angular momentum on the equatorial plane). A picture 
similar to this has been postulated in the two component advective flow model (CT95)
where sub-Keplerian flows flank a Keplerian disc on a equatorial plane. 
However, a complete picture would emerge only after radiative transfer is added.

The authors thank D. Ryu and I. Chattopadhyay for providing helps to include viscosity in the 
original code for non-viscous flow used in Paper I. They are also thankful to an anonymous 
referee for helpful suggestions.

{}

\begin{thebibliography}{}
\def\ref#1\par{\parshape=2 0in 14.5cm 1cm 13.5cm {#1} \par}
\parskip=0pt
\parindent=0pt
\bibitem[]{1} Acheson, D. J., 1990, Elementary Fluid Dynamics, 
Oxford Applied Mathematics and Computing Science Series, Oxford University Press
\bibitem[]{2} Batchelor, G. K., 1967, An Introduction to Fluid Dynamics, Cambridge University Press
\bibitem[]{3} Chakrabarti, S.K. 1989, ApJ 347, 365 (C89)
\bibitem[]{4} Chakrabarti, S. K., 1990, MNRAS 243, 610
\bibitem[]{5} Chakrabarti, S.K. 1993, MNRAS 259, 410
\bibitem[]{6} Chakrabarti, S.K. 1996b, ApJ 471, 237
\bibitem[]{7} Chakrabarti, S. K. 1996b, ApJ 464, 664
\bibitem[]{8} Chakrabarti S. K., Acharyya K. A., Molteni D., 2004, A \& A, 421
\bibitem[]{9} Chakrabarti, S.K. \& Das, S., 2001, MNRAS 327,808
\bibitem[]{10} Chakrabarti, S.K. \& Das, S. 2004, MNRAS 349, 649 
\bibitem[]{11} Chakrabarti, S.K. \&  Molteni, 1993, ApJ 417, 672
\bibitem[]{12} Chakrabarti, S.K. \&  Molteni, D. 1995, MNRAS 272, 80
\bibitem[]{13} Chakrabarti, S. K., \& Manickam, S. G. 2000, ApJ, 531, L41
\bibitem[]{14} Chakrabarti, S.K. \& Titarchuk, L.G. 1995, ApJ 455, 623
\bibitem[]{15} Das, S. \& Chakrabarti, S. K., 2004, IJMPD 13(9), 1955
\bibitem[]{16} Gliozzi, M., Rath, C., Papadakis, I. E., Reig, P., 2010, A\&A 512, 21	
\bibitem[]{17} Giri, K., Chakrabarti, S. K., Samanta, M., \& Ryu, D. 2010, MNRAS 403,516
\bibitem[]{18} Harten, A., 1983, J. Comp. Phys. 49, 357
\bibitem[]{19} Igumenshchev I. V., Chen X., Abramowicz M. A., 1996, MNRAS 278, 236
\bibitem[]{20} Igumenshchev I. V., Abramowicz M. A., 1999, MNRAS 303, 309
\bibitem[]{21} Igumenshchev I. V., Abramowicz M. A., 2000, ApJ 130, 463
\bibitem[]{22} Landau, L. D.; Lifshitz, E. M., Fluid Mechanics (Pergamon Press, Oxford, 1959)
\bibitem[]{23} Lanzafame, G., Molteni, D., \& Chakrabarti, S. K. 1998, MNRAS 299 799
\bibitem[]{24} Lee, S.J, Ryu, D, \& Chottopadhya, I., 2011, ApJ 728, 142
\bibitem[]{25} MacFadyen, A.I. \& Woosley, S.E., 1999, ApJ 524, 262
\bibitem[]{26} Molteni, D., Lanzafame, G., \& Chakrabarti, S.K., 1994, ApJ 425, 161 (MLC94)
\bibitem[]{27} Molteni, D., Ryu, D., \& Chakrabarti, S. K., 1996, ApJ 470, 460
\bibitem[]{28} Molteni, D., Sponholz, H., \& Chakrabarti, S.K., 1996, ApJ 457, 805
\bibitem[]{29} Novikov, I. D. \& Thorne, K. S., 1973, Black Holes, (Eds.), C. DeWitt
\& B. DeWitt, (Gordon and Breach: New York)
\bibitem[]{30} Pringle, J. E., 1981, ARAA 19, 137
\bibitem[]{31} Paczy\'nski, B. \& Wiita, P. J., 1980, A\&A  88, 23 
\bibitem[]{32} Qu, J. L., Lu, F. J., Lu, Y., Song, L. M., Zhang, S., Ding, G. Q., Wang, J.,
 2010, ApJ 710, 836 
\bibitem[]{33} Remillard, R. A., McClintock, J. E., 2006, ARA\&A 44, 49
\bibitem[]{34} Rodriguez, J., Corbel, S., Kalemci, E., Tomsick, J. A. \& Tagger, M.,
2004, ApJ 612, 101            
\bibitem[]{35} Ryu, D., Brown, G. L., Ostriker, J. P. \& Loeb, A., 1995, ApJ 452, 364
\bibitem[]{36} Ryu, D., Chakrabarti, S. K. \& Molteni, D., 1997, ApJ 378, 388
\bibitem[]{37} Smith, D. M., Heindl, W. A., \& Swank, J. H., 2002, ApJ 569, 362
\bibitem[]{38} Shakura N. I., Sunyaev R. A., 1973, A \& A, 24, 337
\end{thebibliography}
\end{document}